\documentclass{article}
\usepackage{jheppub1}
\usepackage{graphicx}
\usepackage{caption}
\usepackage{subfigure}

\title{Using Atom Interferometry to Detect Dark Energy}
 \author{Clare Burrage and }  
 \author{Edmund J. Copeland}
 \affiliation{School of Physics and Astronomy, University of Nottingham, Nottingham, NG7 2RD, United Kingdom}

\emailAdd{clare.burrage@nottingham.ac.uk}
\emailAdd{ed.copeland@nottingham.ac.uk}

\date{today}
\abstract{We review the tantalising prospect that the first evidence for the dark energy driving the observed acceleration of the Universe on giga-parsec scales may be found through metre scale laboratory based atom interferometry experiments. To do that, we first introduce the idea that scalar fields could be responsible for dark energy and show that in order to be compatible with fifth force constraints these fields must have a screening mechanism which hides their effects from us within the solar system. Particular emphasis is placed on one such screening mechanism known as the chameleon effect where the field's mass becomes dependent on the  environment. The way the field behaves in the presence of a spherical source is determined and we then go on to show how in the presence of the kind of high vacuum associated with atom interferometry experiments, and when the test particle is an atom, it is possible to use the associated interference pattern to place constraints on the acceleration due to the fifth force of the chameleon field - this has already been used to rule out large regions of the chameleon parameter space  and maybe one day will be able to detect the force due to the dark energy field in the laboratory. 
  }

\begin{document}
\maketitle
\section{Setting the scene}\label{introduction}
When we go outside on a clear evening and look up in the night sky, then as long as we are not in a major metropolis, we are likely to see a Universe pretty much as Einstein saw it over a century ago when he finally developed the beautiful theory of general relativity \cite{Einstein:1915by}. As far as he was concerned the universe was static, `as required by the fact of the small velocities of the stars' \cite{Einstein1917}. Galaxies or the nearby ones we could see, didn't appear to be doing anything with respect to us, they remained in the same place in the night sky day after day after day. However, there was an unsettling aspect to his newly established theory of gravity, basically it did not care for a static Universe. The simplest solutions which could be thought of as describing the large scale properties of our Universe had it either expanding or contracting, anything but staying where it was. It forced Einstein to consider a radical new term in his equations, a term perfectly acceptable from a mathematical point of view but completely ad hoc as far as physics was concerned. With this extra positive constant contribution to the energy density of the universe Einstein was able to persuade the Universe to remain static, the price paid being that space was curved leading to a finite Universe. This new `Cosmological Constant' acted as `repulsive gravity' balancing the attractive gravity induced by matter. 

 It is somewhat surprising that although Einstein  demonstrated back in 1917 that this `Cosmological Constant' would do the trick,  it seems that it took until 1930 before it was realised that such a static unmoving Universe was unstable \cite{Eddington1930}, a bit like an upright pencil stood on its sharp end is unstable. Just as in that case any small push and the pencil falls to the floor, similarly Eddington showed that any small deviation away from static and the universe would once again either start collapsing or expanding. The need for this Cosmological term vanished as first Slipher \cite{Slipher1917} and then Hubble \cite{Hubble1929} demonstrated that distant galaxies, were actually moving apart from each other with a velocity proportional to their separation. The interpretation was immediately obvious, the space between the galaxies was growing, a bit like the space between polka dots on a party balloon increases as it is blown up. Our Universe was not static, it was dynamical, it was expanding. The story goes that Einstein later commented that the inclusion of this Cosmological Constant in his evolution equations was his `Biggest Blunder' implying that if he had stuck with the original form of the equations he would have predicted the Universe was expanding - quite an achievement ! 
Although he may well have thought it, he should probably have been asking why the Cosmological Constant should be zero, given that it had every right to be in the equations. No one was able to answer this question satisfactorily and yet for decades to come when we considered applying his famous theory to cosmology we would simply discard this term, setting it to zero and arguing there {\it must} be an underlying symmetry present that would cause it to vanish. 

This was the state of play until the mid 1980's around 70 years after the theory first hit the news stands. The Universe's evolution was pretty straightforward to follow even if there remained a few niggling details that required filling in. It began its existence with one of those niggling little details that still remains today, emerging from an initial hot dense region known as the Hot Big Bang,  it went through a period  during which its  evolution  was dominated by the radiation present in the Universe  succeeded by a period during which the evolution was dominated by the matter in the Universe. It was constantly decelerating as might be expected of such a scenario because after all gravity is an attractive force and has a tendency to pull the matter together and slow down the Universe's rate of expansion. 

All seemed consistent, but in the mid 1980s contributions from a number of directions suggested things may not be so simple. Steven Weinberg actually argued that  in order for life as we know it to exist  we should expect to live in a Universe with a Cosmological Constant actually dominating the energy density and suggested it should be of order 60\% of it \cite{Weinberg:1988cp}. In the late 1980s a series of observations looking at the angular correlation of galaxies on the sky showed that they were not described at all well by a Universe dominated by matter, but rather were much better fit by one which was dominated again by a cosmological constant type term \cite{Efstathiou:1990xe}. Strangely enough this key result went by with relatively little comment and it wasn't until the late 1990s that things took a dramatic turn with a series of observations of exploding Supernovas in distant galaxies that led to the three key participants being awarded the 2011 Nobel Prize for Physics, and the necessity for some form of cosmological constant driving the dynamics of the Universe today \cite{Perlmutter:1998np,Riess:1998cb}. Whereas Einstein used the constant in conjunction with a positively curved Universe to make it static, if we abandon the positive curvature condition and allow the Universe to be close to spatially flat today, the cosmological constant will cause it to not only expand but accelerate with distant galaxies receding from each other with ever increasing velocities. Subsequent observations of more supernova and other complementary observations of clusters of galaxies, large scale structure and the small deviations in temperature associated with the otherwise amazingly uniform radiation from the Big Bang have led to a consensus model in which the Universe became dominated around five billion years ago by a source of energy (dubbed dark energy) which at the very least looks like Einstein's famous cosmological constant and is causing the Universe to speed up as time goes on - for a comprehensive review of the current constraints on the cosmological parameters see \cite{Lahav:2014vza,Ade:2015xua}. 

This concordance or vanilla model has a spatially flat universe that is accelerating. It is made up of baryons making up 5\% of the energy density, dark matter (26\%) and dark energy (69\%). Some 13.8 billion years ago the Universe underwent a hot, dense, early phase of expansion yielding both the light elements through nucleosynthesis and the cosmic microwave background (CMB) radiation associated with the primordial photons of that era. Prior to that period it underwent an epoch of accelerated expansion, known as inflation, and during that incredibly short period of expansion time the primordial density perturbations were produced from quantum fluctuations of the scalar inflaton field. These in turn left an imprint on the CMB anisotropy and led to the formation of large-scale structure by simple gravitational instability. This model has been very well tested and seems to fit the data exceptionally well \cite{Ade:2015xua}. 

If we apply Occam's razor, all we need to do is include in Einstein's equations the one term which is allowed but he regretted adding, and we have an explanation for the observed acceleration of the Universe.  However, although the inclusion of that term appears to work, it comes with a huge problem - we live in a world of Quantum Mechanics where small things can have a big impact, and none bigger than here. In fact the cosmological constant problem is the most severe fine tuning problem in physics today.  In the 1960's the great Soviet Physicist Yakov Zel'dovich showed that the cosmological constant was mathematically equivalent to the stress-energy of empty space, meaning that the vacuum could not be dismissed as irrelevant \cite{Sahni:2008zza}. Quantum field theory tells us how quantum mechanics impacts on scalar fields and in particular it tells us that the vacuum state is not empty as we might naively expect, but rather it is filled with virtual particles. The effects of the virtual particles have been measured in their impact on matter fields for example in the shifts of atomic lines and in particle masses. However, here is the problem, when we estimate the energy density associated with the quantum vacuum, these estimates are at best  60 orders of magnitude too large -  something that has become known as the cosmological constant problem. 

There is no convincing solution to the cosmological constant problem although many of us have made attempts to achieve one. What is clear is that it will require us to go beyond the current standard models of cosmology and particle physics with the solution being one that must interact with both matter and gravitational fields.    Modifying our physical theory will generally introduce new scalar fields, and searching for these fields presents an important opportunity to learn about the solution to the cosmological constant problem.  These scalar  fields may drive the acceleration of the expansion of the universe directly, or they may be a subdominant component of our current universe.   This new scalar degree of freedom needs to very light, so that the Compton wavelength of the field can be comparable to the size of the universe today,  and 
will couple to matter fields. Any such field that drives the current acceleration of the Universe is known as dark energy - for a review see \cite{Copeland:2006wr}.  It comes in different guises, it could be a new particle in its own right, or a component of a massive graviton. It could arise from a string compactification or from another source entirely.  It can also arise because  general relativity does not hold on cosmological scales and has to be replaced by a more complete theory of gravity - for a review see \cite{Clifton:2011jh}. Some of these modified gravity theories behave as if they are in the class of chameleon models we will be discussing in this article in that they can lead to density or curvature dependent effects as chameleon fields do. 
  Whatever the source of the field, the presence of  light scalar degrees of freedom generally poses a problem.  Their coupling to matter means that they will mediate long range fifth forces that have not yet been detected on Earth or in the solar system \cite{Adelberger:2003zx}. Now one way of alleviating this tension between theory and observation is through the introduction of screening mechanisms, which allow the properties of the  field, and the force that it mediates, to vary depending on the environment at the cost of making the scalar field theory non-linear.  
	
There are a number of scalar theories admitting screening behaviour that have been constructed.  These include the Chameleon \cite{Khoury:2003aq,Brax:2004qh}, Dilaton \cite{Brax:2011ja} and the Symmetron \cite{Hinterbichler:2010es,Olive:2007aj,Pietroni:2005pv} models where non-linearities in the scalar potential or form of the coupling to matter result in the mass of the field, or the strength of the coupling becoming environment dependent. This means that for suitable parameter choices the worrisome scalar force is suppressed in regions of higher density including those used in experimental searches for fifth forces,  screening them from detection with current experiments.  However, despite being explicitly designed to evade the constraints of current fifth force searches, a benefit of these theories is that alternative scenarios can be devised to search for the existence of such scalar fields in laboratory based experiments. The reason is that in a laboratory vacuum the extremely low density ensures that sufficiently small objects are not screened from the scalar field and are thus sensitive probes of dark energy. In this particular unscreened regime the force sourced by the dark energy scalar could significantly exceed the gravitational interaction. Yet no deviation from general relativity would be seen with larger sources or in less diffuse environments. 

 There have been a number of ways of probing the chameleon screening mechanism in the laboratory proposed or implemented in recent years.  These include neutron spectroscopy experiments conducted at an energy scale of $10^{-14}$ eV, that constrain departures from Newtonian gravity over micron distances and can be used to  introduce new bounds on the parameter space of the chameleon model \cite{Jenke:2014yel,Lemmel:2015kwa}. There are also Casimir-like experiments under way to study the chameleon field over very short distance scales \cite{Brax:2010xx,Brax:2014zta}. Part of the remaining parameter space has now been  probed with  ultra-cold atom interferometry experiments and in the near future much of the remaining space will be covered by experiments that  should be able to detect even a chameleon field with Planck suppressed couplings \cite{Burrage:2014oza,Hamilton:2015zga}. 
In this article we will review the chameleon mechanism and attempts to search for it using atom interferometry.

\section{Screening mechanisms}\label{screening}
Any scalar field $\phi$ which is a candidate for dark energy necessarily implies that we are introducing  new degrees of freedom into our system. Whatever it is it must satisfy two particularly severe constraints. First, in order to be useful as a dark energy candidate 
it must have a mass no larger than the current Hubble parameter, $m_{\phi} \leq H_0 \sim 10^{-30}\mbox{ meV}$, otherwise it would play no role in the low energy dynamics of the Universe. The second constraint arises because $\phi$ has to couple to standard model particles, it has no choice since the standard model fields themselves contribute to the energy density of the vacuum with an amount at least ${\cal O}(\rm TeV)^4$.  Therefore any scalar fields that arise as part of a solution to the cosmological constant problem must interact with both gravitational and Standard Model fields. The consequence is that the $\phi$ field mediates an interaction or force between standard model fields, the range of that force being $\sim m_{\phi}^{-1}$. In other words this new force is mediated across the observable universe $\sim H_0^{-1}$. 

Why have we not seen evidence of this extra fifth force? Such forces have been searched for in particular in the solar system and the constraints on their magnitude are impressive, basically we see no evidence for them \cite{Adelberger:2003zx}. On the other hand, tests on cosmological scales are far less restrictive. It opens up the possibility that in order to account for dark energy, we might be seeing a fifth force  operating on cosmological scales, but that it is somehow being screened from us on solar system and laboratory scales. There are three key {\it screening mechanisms} that have been proposed, and they rely on the fact that locally (i.e. within the solar system for example), the density of matter is high compared to cosmological scales, and it is this property that suppresses the deviations from general relativity. It makes sense, the mean cosmological density is something like $\sim10^{-29} {\rm gcm^{-3}}$ whereas the density of the solar system is $\sim 10^{-22}{\rm gcm^{-3}}$ and planets are $\sim {\rm gcm^{-3}}$. As we will see shortly, given that the solar system is over a million times more dense than the cosmological background, it is possible to develop mechanisms which will allow us to screen this fifth force in dense environments like the solar system. What will be extraordinary though, and is the purpose of this article, is that within the solar system it will be possible to set up situations where locally the force fails to be screened because we can set up a scenario involving an extraordinary vacuum and a screening object that is just not dense enough to react to the force and suppress it. Such objects are atoms, and such situations are found in the vacuum of atom interferometry experiments \cite{Burrage:2014oza}.  

It is worth spending a little while explaining the nature of the screening mechanisms. A very nice discussion of this can be found in Joyce et al \cite{Joyce:2014kja} and we will follow their basic approach here. When considering the dynamics of a physical system we usually consider the Lagrangian, which is effectively the difference between the kinetic energy and potential energy of the system. For the case of a general scalar field $\phi$ which is conformally coupled to matter then the Lagrangian can be written as 
\begin{equation} \label{gen-lag}
{\cal L} = - \frac{1}{2} Z^{\mu \nu}(\phi, \partial \phi,...) \partial_\mu \phi  \partial_\nu \phi - V(\phi) + g(\phi) T^{\mu}_{\mu} 
\end{equation} 
where the indices $\mu,\nu =0,1,2,3$ refer to the four spacetime coordinates, $Z^{\mu \nu}$ is a way of encoding derivative self-interactions of the field (often called non-canonical kinetic terms), $V(\phi)$ is the potential associated with the field, $g(\phi)$ is the coupling strength of the field to matter and $T^{\mu}_{\mu}$ is the trace of the matter stress-energy tensor. If the source is non-relativistic, such as ordinary baryonic matter for example, then the pressure vanishes and we have $T^{\mu}_{\mu} = -\rho$. To a first approximation we can consider the sources as being point like with a constant density ${\cal M}$, hence $\rho = {\cal M} \delta^3 (\vec{x})$. If we then expand the field about its background solution $\bar{\phi}$ as $\phi = \bar{\phi} + \varphi$, the equation of motion for the perturbation $\varphi$ follows from the Euler-Lagrange equations as  
\begin{equation} \label{EL-eqn}
Z(\bar{\phi}) \left(\ddot{\varphi} - c_s^2(\bar{\phi}) \nabla^2 \varphi\right) + m^2(\bar{\phi}) \varphi = g(\bar{\phi}) {\cal M} \delta^3 (\vec{x})
\end{equation}
where $\ddot{\varphi} \equiv \frac{\partial^2 \varphi}{\partial t^2}$, $Z(\bar{\phi}) = Z^\mu_\mu(\bar{\phi})$, $c_s^2(\bar{\phi})= Z_{ii}(\bar{\phi})/Z(\bar{\phi})$ is an effective sound speed, $m^2(\bar{\phi}) \equiv  \frac{d^2 V}{d\phi^2}|_{\bar{\phi}}$ is the mass of the fluctuating field. The background value $\bar{\phi}$ is considered as being set by the other background quantities such as the local energy density $\bar{\rho}$. If we assume it is homogeneous over the scales of interest then we can write down a resulting static potential 
\begin{equation}\label{static-pot}
V(r) = - \frac{g^2(\bar{\phi})}{Z(\bar{\phi}) c_s^2(\bar{\phi})} \frac{\cal M}{ 4\pi r}  \exp\left(- \frac{m(\bar{\phi}) r}{ \sqrt{Z(\bar{\phi})} c_s(\bar{\phi})}\right)
\end{equation}
 When $m(\bar{\phi})=0$, and the force mediator is massless, the potential in  Equation (\ref{static-pot}) has the $1/r$ form familiar from Newtonian gravity.
The force experienced by a test particle moving in this potential is
\begin{equation}\label{force-eqn}
F(r) = - {dV(r) \over dr} 
\end{equation} 
and it follows from (\ref{static-pot}) that the corresponding force is attractive\footnote{We have assumed here that $Z(\bar{\phi})$ is positive.  If this assumption is not made then the whole theory becomes unstable.}. The problem we face with light scalar fields mediating forces now becomes clear. Assuming the other parameters have natural values ${\cal O}(1)$ then it follows that the perturbation $\varphi$ mediates a gravitational strength long range force $F_\varphi \sim 1/r^2$, something which is not allowed by solar system tests of general relativity. Therefore if we are to have a light scalar field present yet be compatible with fifth force tests, how can we do it? The three known answers can be found in (\ref{static-pot}). The couplings $g, Z, c_s$ and $m$ in (\ref{static-pot}) are all background dependent quantities, relying on the value of $\bar{\phi}$. Given that $\bar{\phi}$ depends on the background energy density $\bar{\rho}$, we see that the required screening mechanisms arise from making the parameters depend on the environment. The three ways they manifest themselves are through \cite{Joyce:2014kja}:
\begin{itemize}
\item Weak Coupling - the coupling to matter $g$ depends on the environment, so that in high density regions, such as the solar system where we perform local tests of gravity, the coupling is very small leading to a fifth force which is so weak it satisfies the constraints. This is simply because from (\ref{force-eqn}) the force $F \propto g^2$. In contrast in regions of low density such as on cosmological scales, $g \sim {\cal O}(1)$ resulting in a fifth force of gravitational strength and leading to the acceleration we observe on those scales. Examples of this include the symmetron or varying-dilaton theories. 
\item Large inertia - In this case the kinetic function $Z(\bar{\phi})$ depends on the environment so that its second derivative becomes large below a given length scale. This is known as Vainshtein screening.
\item Large mass - From the exponential dependence in (\ref{static-pot}) it is clear that another possibility is to allow the mass of fluctuations $m(\bar{\phi})$ to depend on the background matter density. The effect then is that in regions of high density such as the solar system, the field acquires a large mass, making its effects short range and so unobservable. However on cosmological scales where the ambient density is very low, the scalar becomes light and can once again mediate a fifth force of gravitational strength. This type of screening is of the chameleon type and testing for its presence in nature is the main focus of this article.
\end{itemize}   

\section{The chameleon mechanism}\label{chameleon}
\begin{figure}
\begin{center}
\begin{minipage}{110mm}
\subfigure[Low density.]{
\includegraphics[width=0.5\textwidth]{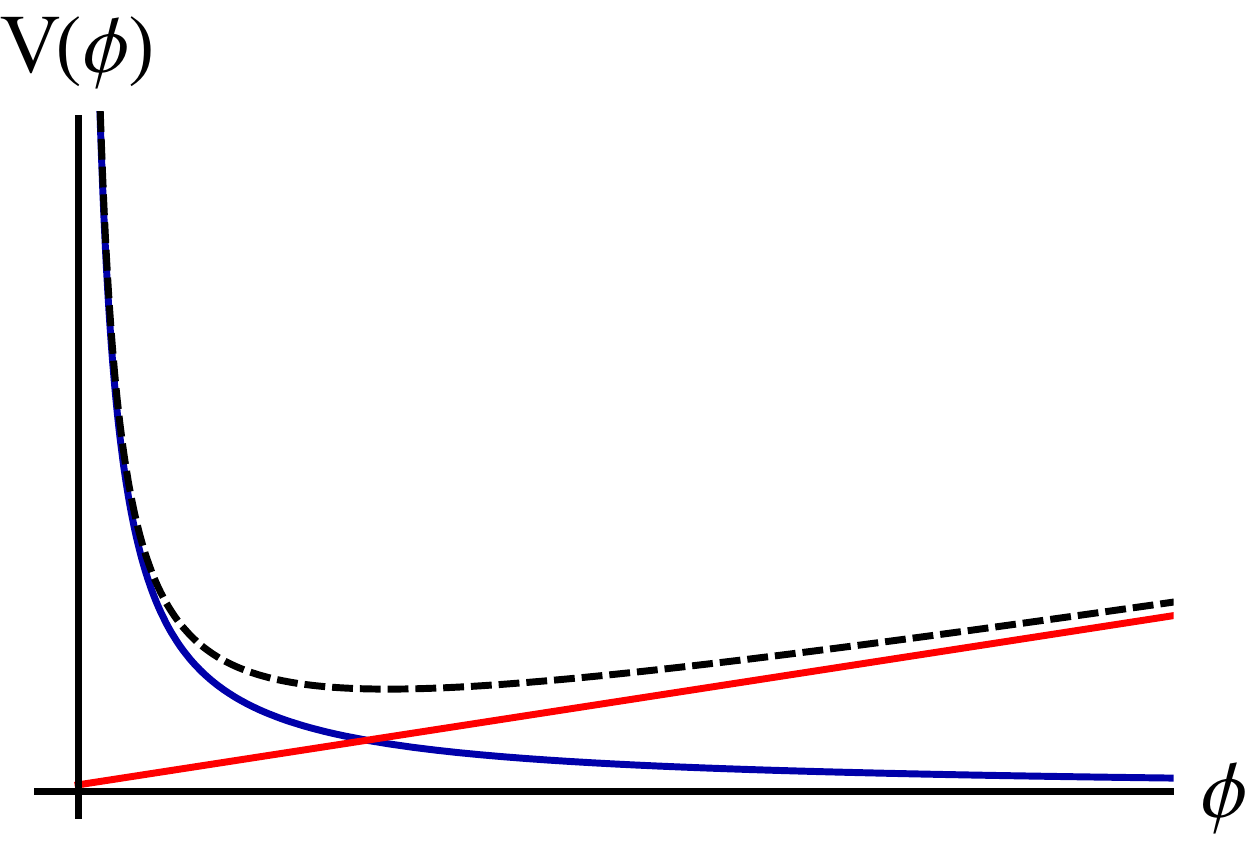}}
\subfigure[High density.]{
\includegraphics[width=0.5\textwidth]{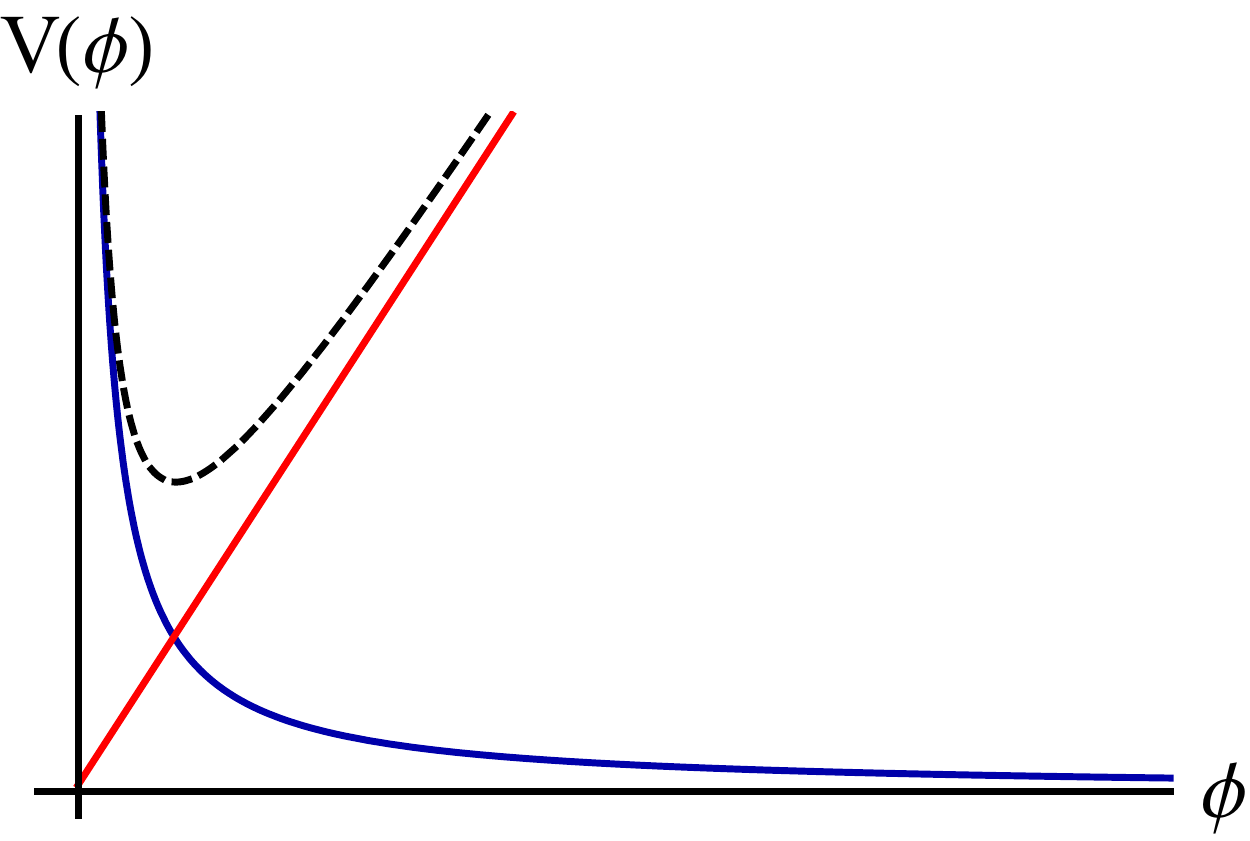}}
\caption{Sketch of the chameleon effective potential.  The blue line indicates the bare potential, the red line the contribution from the coupling to matter, and the black dashed line the sum of the two contributions.} \label{fig:potential}
\end{minipage}
\end{center}
\end{figure}

It is time to explain what the chameleon field is and how it changes depending on the local environment it experiences  in order to camouflage itself and evade detection (hence chameleon of course!) \cite{Khoury:2003aq,Khoury:2003rn}. The source which determines the local density will be considered as being static and spherical for simplicity although allowing it to be of varying shapes can be a very useful additional degree of freedom \cite{Burrage:2014daa}. 
The chameleon is a scalar field, $\phi$, whose behaviour is determined by the following action:
\begin{eqnarray}
S&=&\int d^4x\;\sqrt{-g}\left[\frac{1}{16 \pi G}R -\frac{1}{2} \nabla_{\mu} \phi \nabla^{\mu} \phi -V(\phi)\right] \nonumber \\
& &+\int d^4 x \; \mathcal{L}_{(m)} (\psi_{(m)}, \Omega^{-2}(\phi)g_{\mu\nu})\;,
\label{eq:chamaction}
\end{eqnarray}
where $g_{\mu\nu}$ is the space-time metric, $g$ is the determinant of the metric and $R$ the associated Ricci curvature.  $V(\phi)$ is the chameleon potential
 and $S_{(m)}=\int d^4 x \; \mathcal{L}_{(m)} (\psi_{(m)}, \Omega^{-2}(\phi)g_{\mu\nu})$ is the matter action.  Matter fields, $\psi_{(m)}$ move on geodesics of the conformally rescaled metric $\tilde{g}_{\mu\nu}=\Omega^{-2}(\phi)g_{\mu\nu}$  and the function $\Omega(\phi)$ determines the coupling between the scalar and matter fields. A typical coupling is of the form $\Omega(\phi) = e^{-\phi/2M}$ where $M$ is a constant energy scale governing the strength of the coupling. 
For the situations considered in this article it is sufficient to approximate matter distributions as perfect fluids with  density $\rho$ and pressure $p$. 
Also, for the cases of interest to us the value of the field will be such that $\phi / M \ll 1$. Given that we can then Taylor expand the coupling function $\Omega$ around $\phi = 0$ and only keep the first term in the series that is relevant in the equation of motion leading to Equation (\ref{eq:eomspsym}), which can be interpreted as the chameleon moving in a density-dependent potential:
\begin{equation}
V_{\rm eff}(\phi) =V(\phi) +\left(1 +\frac{\phi }{M}\right)\rho\;.
\label{eq:Veff}
\end{equation}
The scale $M$ is constrained, by precision measurements of atomic structure and searches for fifth forces \cite{Mota:2006fz,Brax:2010gp,Upadhye:2012qu}, to lie in the range $10^{10}\mbox{ TeV} \leq M \leq M_P \sim 10^{15} \mbox{ TeV}$. For a static, spherically symmetric configuration sourced by non-relativistic matter the equation of motion for $\phi$ is:
\begin{equation}
\frac{1}{r^2}\frac{d}{dr}\left[ r^2 \frac{d \phi(r)}{dr} \right] = \frac{d}{d\phi}V_{\rm eff}(\phi) = \frac{d}{d\phi}V(\phi) + \frac{\rho}{M}\;,
\label{eq:eomspsym}
\end{equation}
For simplicity we specialise to a common choice of the bare chameleon potential,
 $V(\phi)=\Lambda^5/\phi$, where $\Lambda$ is a constant energy scale. Note as mentioned earlier, the non-linear nature of the potential. The minimum of the corresponding effective potential, and the mass of fluctuations around this minimum follow:
\begin{eqnarray}
\phi_{\rm min}(\rho)&=&\left(\frac{\Lambda^5 M}{\rho}\right)^{1/2}\;, \label{eq:phimin}\\
m_{\rm min}(\rho) \equiv \sqrt{\left.{d^2 V_{\rm eff} \over d\phi^2} \right|_{\phi_{\rm min}}} &=& \sqrt{2}\left(\frac{\rho^3}{\Lambda^5 M^3}\right)^{1/4}\;.\label{eq:mmin}
\end{eqnarray}
Some key density dependent properties of the chameleon field follow from Equations (\ref{eq:phimin}) and (\ref{eq:mmin}). For large densities $\rho$, we see that the field tries to reach as small a value as possible with a corresponding large mass, whereas for low densities the field value increases and the mass of the field decreases accordingly, as we expect for a screening mechanism. This behaviour can be seen in Figure \ref{fig:potential}.

Given the fact that the sources for the chameleon field we will be studying are spherically symmetric and of constant density, in the chameleon equation of motion the source term is 
\begin{equation}
\rho(r)=\rho_A \Theta(R_A-r)+\rho_{\rm bg}\Theta(r-R_A)\;,
\label{eq:sphereom}
\end{equation}
where $\rho_A$ and $R_A$ are respectively the  density and radius of the source  (which therefore has mass $M_A = (4/3)\pi \rho_A R_A^3$).  In addition, $\Theta(x)$ is the Heaviside step function, and $\rho_{\rm bg}$ is the density of the background environment surrounding the ball, which to a first approximation can be assumed to be constant density and extends to infinity.

It is possible to solve the equation of motion for the chameleon in a piecewise manner, by making  approximations to the chameleon effective potential. The reader not too concerned with the technical details may wish to skip the remainder of this section, but it is enlightening to see how the field reacts to the source. Far away from the source, for example on cosmological scales, the scalar field will be close to its background value $\phi_{\rm bg}$ which recall is given by Equation (\ref{eq:phimin}) with $\rho=\rho_{\rm bg}$. The contribution of the effective potential to the equation of motion is then well approximated by the mass term arising from a harmonic expansion of the potential $V_{\rm eff}(\phi) = V_{\rm eff}(\phi_{\rm bg}) + \frac{m_{\rm bg}^2}{2} [\phi - \phi_{\rm bg}]^2 + ...$
\begin{equation}
\frac{1}{r^2}\frac{\partial}{\partial r}\left(r^2 \frac{\partial \phi}{\partial r}\right)=m_{\rm bg}^2(\rho_{\rm bg})[\phi-\phi_{\rm bg}]\;,
\end{equation}
where $m_{\rm bg}^2 = d^2 V_{\rm eff}/d\phi^2|_{\phi_{\rm bg}}$.  Solutions to the equation of motion are:
\begin{equation}
\phi =\phi_{\rm bg} +\frac{\alpha}{r}e^{-m_{\rm bg}r} +\frac{\beta}{r}e^{m_{\rm bg}r}\;,
\label{eq:out}
\end{equation}
and the field profile must decay at infinity, implying $\beta=0$.

Inside the source ball, the high density $\rho_A$ has the effect of moving the minimum of the effective potential to a lower field value $\phi_A \equiv \phi_{\rm min}(\rho_A) < \phi_{\rm bg}$. 
We now have two possible types of solution in this region. In the first case, the field $\phi$ decreases inside the ball, but remains everywhere greater than $\phi_A$, a regime we call weakly perturbing. The solutions for $\phi$ and $d\phi/dr$  must match at $r=R_A$, and it turns out that the consistent solution which does just that in this weakly perturbing regime is
\begin{equation}
\phi=\phi_{\rm bg}-\frac{1}{8 \pi R_A}\frac{M_A}{ M}\left\{\begin{array}{lc}
3-\frac{r^2}{R_A^2}\;, & r<R_A\;,\\
2\frac{R_A}{r}e^{-m_{\rm bg}r}\;, & r>R_A\;.
\end{array}\right.
\end{equation}
Both of these expressions have been simplified by taking $m_{\rm bg}R_A\ll 1$.  For the experiments we are considering here, $R_A \sim 1\mbox{ cm}$ and $\rho_{\rm bg}$ corresponds to a good vacuum, making this approximation valid over almost all the relevant values of the parameters $\Lambda$ and $M$.  The weak perturbation is valid in the domain:
\begin{equation}
\frac{1}{4 \pi R_A}\frac{M_A}{M}\ll\phi_{\rm bg}\;.
\end{equation}

The second type of solution, is called strongly perturbing, and it corresponds to the case where the field inside the ball actually reaches $\phi_A$.  If it happens it will be near the centre, let us say within a radius $S$. Once again making sure the values of $\phi$ and $d\phi/dr$ match both at $r=S$ and $r=R_A$ it can be shown that the unique solution is \cite{Burrage:2014oza}
\begin{equation}
\phi=\left\{\begin{array}{lc}
\phi_A\;,  & r<S\;,\\
\phi_A +\frac{1}{8\pi R_A}\frac{M_A}{ M }\frac{r^3-3S^2 r +2S^3}{rR_A^2}\;, & S < r<R_A \;,\\
\phi_{\rm bg}-\frac{1}{4 \pi R_A}\frac{M_A}{M}\left(1-\left(\frac{S}{R_A}\right)^3\right)\frac{R_A}{r}e^{-m_{\rm bg}r}\;, & R_A<r\;,
\end{array}\right.
\label{eq:thinshellprofile}
\end{equation}
where
\begin{equation}
S=R_A\sqrt{1-\frac{8\pi}{3}\frac{M}{M_A}R_A \phi_{\rm bg}}\;.
\end{equation}
Once again we have had to make approximations, two in this case:  The first is $m_{\rm bg}R_A \ll1$, the same approximation that we made in the case of the weakly perturbing ball.  The second is $\phi_{\rm bg} \gg \phi_A$, which is well justified here because we are considering a ball of solid material surrounded by a vacuum.  The scalar field has the strongly perturbed profile provided $0\leq S\leq R_A$, which is equivalent to
\begin{equation}
\frac{3}{8\pi R_A}\frac{M_A}{M}\geq \phi_{\rm bg}\;.
\end{equation}
Khoury and Weltman\cite{Khoury:2003rn} called this the thin-shell regime  because the value of the scalar field drops from $\phi_{\rm bg}$ to $\phi_A$ over a thin region near the surface of the ball.

Back to the physics. It proves convenient to write the scalar field outside the ball in a universal form for both weakly and strongly perturbing objects:
\begin{equation}
\phi=\phi_{\rm bg}- \lambda_A \frac{1}{4 \pi R_A}\frac{M_A}{M} \frac{R_A}{r} e^{-m_{\rm bg}r}
\label{eq:universal}
\end{equation}
where
\begin{equation}
\lambda_{A} = \left\{ \begin{array}{lc}
1\;, & \rho_A R_A^2< 3 M \phi_{\rm bg}\;,\\
1-\frac{S^3}{R_A^3}\approx \frac{3M\phi_{\rm bg}}{\rho_A R_A^2}\;, & \rho_A R_A^2> 3 M \phi_{\rm bg}\;.
\end{array}\right.
\label{eq:lambdaA}
\end{equation}
The parameter $\lambda_A$ determines how responsive the chameleon field is to the object. The effect of the chameleon field is that it pulls a point test particle towards the spherical test mass with acceleration
\begin{equation}
a_{\phi}=\frac{1}{M}\partial_{r}\phi\;.
\label{eq:accn}
\end{equation}
This may be compared with the usual (Newtonian) gravitational acceleration, $a_N = G M_A /r^2$. At the distances of interest here, $m_{\rm bg}r\ll1$,  the ratio is
\begin{equation}
\frac{a_{\phi}}{a_N}=\frac{\partial_{r}\phi}{M}\frac{r^2}{GM_A}=2 \lambda_A \left(\frac{M_P}{M}\right)^2\;,
\end{equation}
 $M_P$ is the reduced Planck mass: $M_P^2 =1/(8 \pi G)$. Since $ \left(\frac{M_P}{M}\right)^2$ is somewhere in the range $1 - 10^{28}$, there is every possibility that the chameleon force on a test mass can greatly exceed the Newtonian force, except in cases when $\lambda_A$ is exceedingly small. For most test objects of interest to us, it is indeed the case that $\lambda_A\ll 1$, but as we will see for the case of atoms we can find ourselves in a regime where it is not small and hence can lead to an appreciable effect as the chameleon force fails to be screened from us.

For two source masses, $A$ and $B$, the total gravitational and chameleon  force between the spherical sources is
\begin{equation}
F_r = \frac{G M_A M_B}{r^2} \left[ 1 + 2 \lambda_A \lambda_B\left(\frac{M_P}{M}\right)^2\right]
\end{equation}
where the subscripts $A$ and $B$ label the two objects and the $r$ subscript on $F_r$ indicates that this is the force  along the line connecting the centres of the two spheres.  $M_{A,B}$ are the masses of the two spheres and the $\lambda_i$ are defined in Equation (\ref{eq:lambdaA}).
 For planets, stars and laboratory test masses we find that the corresponding $\lambda_i$ is much smaller than one, and so the chameleon contribution to the total  force between the two spheres is suppressed.  However in a laboratory vacuum atoms can have $\lambda_{\rm atom}$ of order one.  This gives any measurement of forces that is performed with atoms, rather than larger test masses, an advantage because there is no suppression due to the $\lambda$ factor.

\section{Atom interferometry}

\begin{figure}[tp]
\centering
\includegraphics[scale=1]{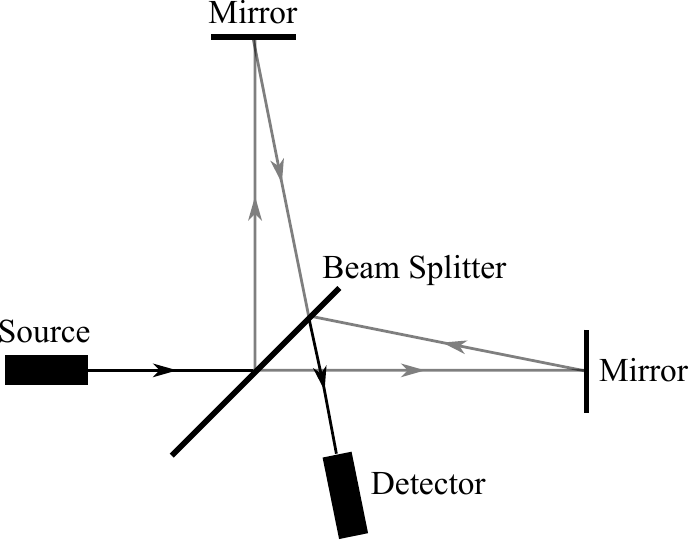}
\caption{Michelson interferometer.   }
\label{fig:Minter}
\end{figure}

Interferometry is a common technique in physics.  A wave is split into two parts and then recombined to give an interference pattern in order to learn about the properties of the waves or the paths that they have travelled on.  The Michelson interferometer, shown in Figure \ref{fig:Minter} interferes light rays that travel along two separate paths.  If there are no physical differences between the paths the light will return in phase, however if there are differences - for example if one of the arms is longer than the other - then the light rays will be out of phase when they are recombined, by an amount that depends on the difference between the two paths.  The Michelson interferometer was famously used by Michelson and Morley to prove that light does not propagate through an ether \cite{MM}.

Atom interferometry relies on the same principles as a Michelson interferometer, a wave is split into two parts that travel along different paths, and are then recombined.  The  difference is that the wave is made of atoms and not light.  It is therefore an intrinsically quantum mechanical experiment, relying on the concept of wave-particle duality.  Any forces that act on the atoms while they are propagating along the arms of the interferometer will modify the properties of the wave and result in an interference pattern.
Atom interferometry experiments with the ability to detect gravitational strength forces were first performed by Peters, Chung and Chu at Stanford University \cite{0026-1394-38-1-4}.  They were able to measure the local acceleration due to gravity with an accuracy of $\Delta g/g = 2 \times 10^{-8}$ using Caesium atoms.  In what follows we describe the theory underlying an atom-interferometry experiment, and we refer those readers interested in the practical details of performing such an experiment to \cite{0026-1394-38-1-4}. It was later realised that atom-interferometry could be used not just to measure the Newtonian gravitational force, but with sufficiently precise measurements it could search for fifth forces coming from beyond standard model physics \cite{Dimopoulos:2003mw} and  it could also be used to test  general relativity \cite{Dimopoulos:2008hx}.  The ability of atom interferometry to constrain beyond the standard model physics is normally due to the unprecedented precision achievable with this technique.  In contrast,  here we present an experimental approach to testing theories of dark energy which relies on unprecedented sensitivity because atoms are so much smaller than other objects previously used to search for new forces. 

\begin{figure}[tp]
\centering
\includegraphics[scale=1]{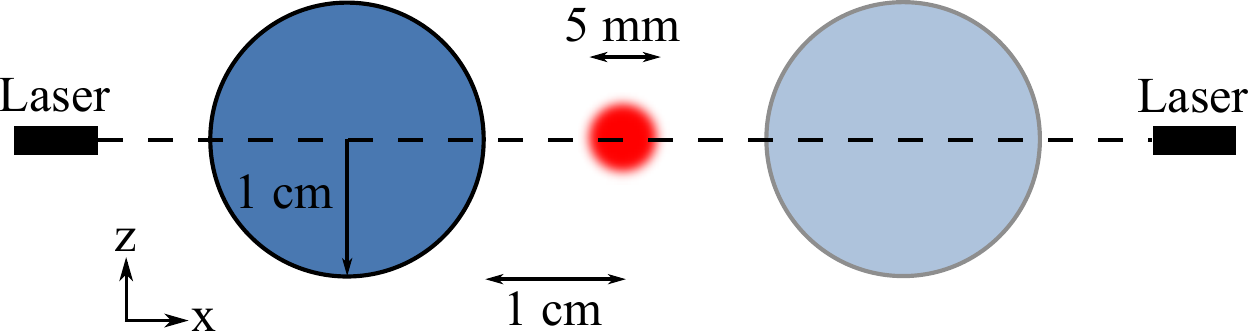}
\caption{Outline of the proposed experimental set up.  The  atoms move freely within the red region at the centre of the Figure. The source mass, indicated by the blue circle, is then moved from its initial position on one side of the cloud of atoms, to its mirror image, indicated by the shaded blue circle and the experiment is repeated.  The black dashed line indicates the direction of propagation of the laser beams. }
\label{fig:drawing}
\end{figure}

A rough sketch of the experiment is shown in Figure \ref{fig:drawing}.  A cloud of atoms is launched in a fountain in the vicinity of a macroscopic spherical mass which is the source of the chameleon force acting on the atoms.  The Figure indicates typical distance scales for such an experiment.

\subsection{Manipulating atoms}
\begin{figure}[tp]
\centering
\includegraphics[]{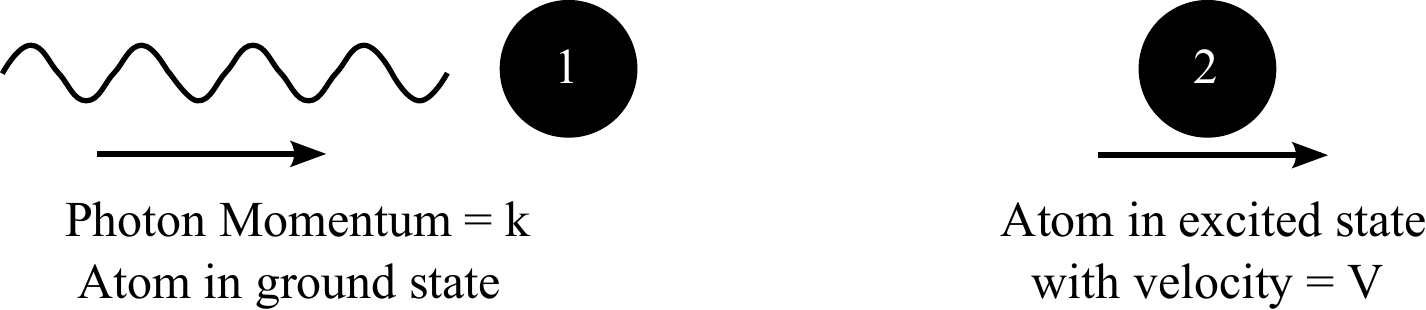}
\caption{An atom inherits momentum from an absorbed photon. }
\label{fig:atoms}
\end{figure}

Let us start by describing how atoms are moved around inside an experiment.  We assume that the atoms we are working with are very cold, so that we can neglect their thermal motion, and consider that we start out with a single stationary atom. This atom has two energy levels, $E_1$ and $E_2$, and the atom is initially at rest in the ground state $E_1$.   We now shine a laser beam at the atom and the frequency of the laser is tuned so that the energy of each photon is exactly $E_2-E_1$.  If the atom absorbs the photon it is excited into state 2.  In order to conserve momentum, the atom must also  have picked up the momentum that was originally carried by the incoming photon.  The excited atom will therefore have a velocity $V= k/M$ where $k$ is the momentum of the incoming photon and $M$ the mass of the atom.  This is shown in Figure \ref{fig:atoms}.

\begin{figure}[tp]
\centering
\includegraphics[scale=0.7]{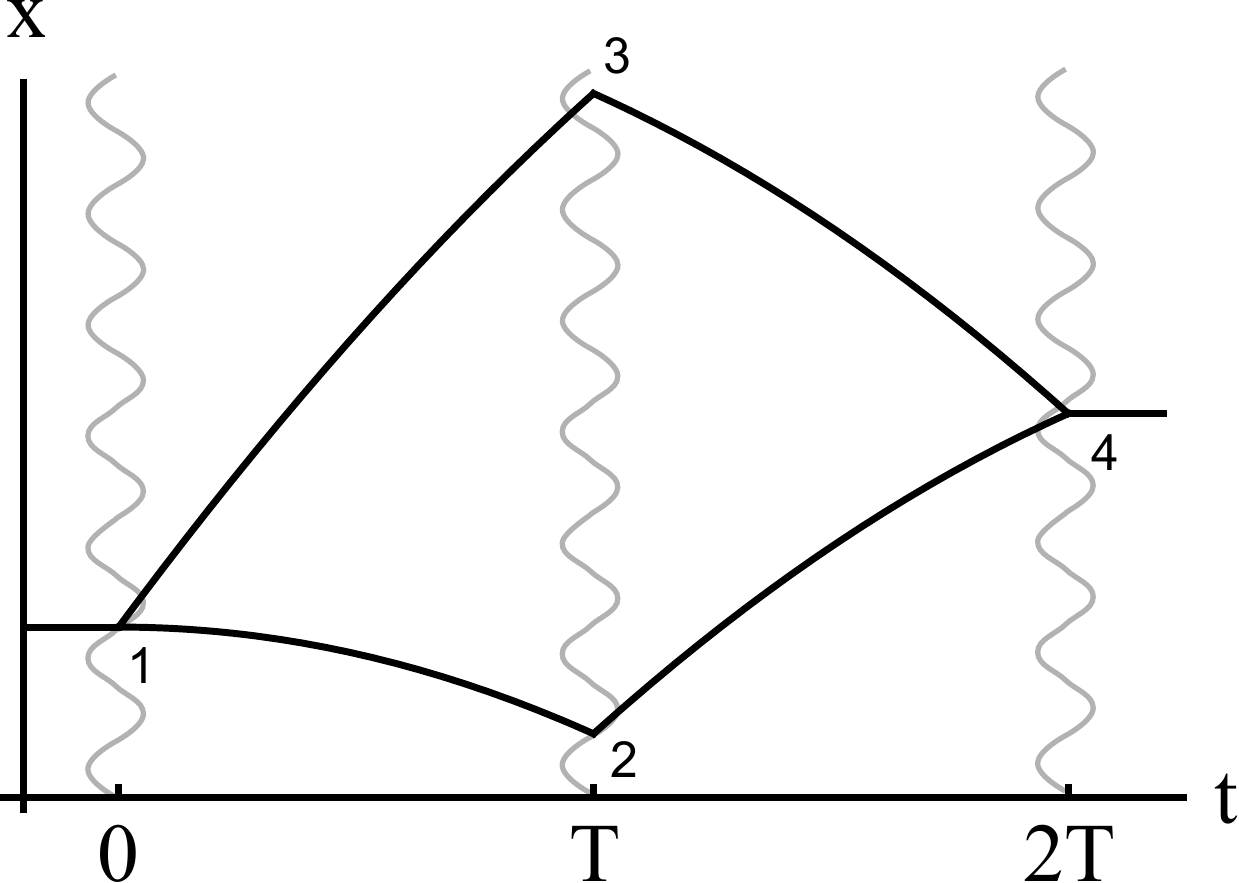}
\caption{Sketch of an atom interferometer.  Interactions between counter-propagating laser beams (grey lines) and atoms (black lines) can be used to give momentum to atoms and put them into a superposition of states which travel along the two arms of the interferometer.  A sequence of three pulses, separated by time $T$, is needed to split and recombine the atomic wave-function. $k_1$ and $k_2$ are the wave-numbers of the laser beams. A chameleon field gradient in the $x$ direction curves the trajectories of the atoms, and this determines  the probability of observing the atom to be in a given state at the output of the interferometer.  }
\label{fig:interferometer}
\end{figure}

If the system is not observed at this point, we do not know whether it has absorbed the photon or not, therefore it is in a superposition of two states; the first where the atom is still stationary and in its ground state, and the second where the atom has been excited and is moving with velocity $V$.  An atom interferometry experiment repeats this process three times to  put an atom into a superposition of states that travel on two different paths shown in Figure \ref{fig:interferometer}.  In between the numbered points the atoms fall freely, and at the numbered points there is a possibility that the atoms interact with a laser beam, which changes the velocity of the atom.  
As we have seen the atom is given momentum when it absorbs a photon, and gets excited from its ground state. The same process happens in reverse when the atom loses energy by a process of stimulated emission.  The presence of the laser beams encourages an excited atom to emit a photon in the direction of the laser beams and to decay back down into its ground state.  Again conservation of momentum means that as the atom decays into the ground state it transfers some of its momentum to the photon, and so the atom's velocity decreases accordingly. 

\subsection{Probability in quantum mechanics}
Atom interferometry  relies on the  interference between the wave description of the atom travelling along two possible paths.  In  the Michelson interferometer in Figure \ref{fig:Minter} the light beam is split into two parts that travel along the two arms of the interferometer.  In an atom interferometer, the principle of quantum superposition allows one atom to explore both possible paths simultaneously.   We will now show that the amount of quantum interference between these two possible paths is sensitive to the forces acting on the atoms. To do this we first review how the probabilities of events are combined in quantum mechanics. 

The interpretation of quantum mechanics as a theory in which particles explore all possible paths between two points of observation is  due to Feynman \cite{feynman} and based on earlier ideas of Dirac \cite{dirac}.  If a particle is observed to be at point A, the probability of finding it after some time $t$ at point B  is found by considering all possible paths that the particle could have explored between A and B.
This probability  is  $P(B,A)=|K(B,A)|^2$ where  $K(B,A)$ is called the kernel 
\begin{equation}
K(B,A) =\sum_{\rm paths} \theta[x(t)]
\end{equation}
where $\theta [x(t)]$ can be thought of as the amplitude for the given path x(t). The sum is performed over all paths $x(t)$ which join the points A and B. The contribution of each path has a phase proportional to the classical action $S$
\begin{equation}
\theta[x(t)] = C e^{(i/\hbar) S[x(t)]}
\end{equation}
where the pre-factor $C$ is just a normalisation constant. 
This is known as the path integral formalism of quantum mechanics; readers interested in the details of this formalism, and how it connects to the  Schr\"{o}dinger description of quantum mechanics are referred to the excellent book by Feynman and Hibbs \cite{feynman}. 

To see how this differs from the notion of  probability in classical mechanics we consider a useful  example. If there are only two possible routes that can be taken between point A and point B, then we can write the probability that the particle travels on path 1  as $P_1 = |\theta_1|^2$, and the probability that the particle travels on path 2 as $P_2=|\theta_2|^2$. But the probability that the particle arrives at point B after travelling on either path 1 or path 2 is 
\begin{equation}
P_{12} = |\theta_1 + \theta_2|^2.
\end{equation}
If we write $\theta_1 = r_1 e^{i\varphi_1}$ and $\theta_2 = r_2 e^{i\varphi_2}$, so that $P_1=r_1^2$ and $P_2=r_2^2$ then we find
\begin{align}
P_{12} &= |r_1 e^{i\varphi_1}+r_2 e^{i\varphi_2}|^2\\
&= P_1 + P_2 + 2 r_1 r_2 \cos(\varphi_1 -\varphi_2)
\end{align}
So if the phase accumulated along the two paths  differs by an amount that is not an odd integer multiple of $\pi/2$, then we will find that in quantum mechanics probabilities do not add in the familiar way $P_{12} \neq P_1 +P_2$.  This is why we say that the two possible paths interfere with one another. A very similar expression appears in the double slit experiment that is used to show that particles also have a wave like nature.  The probability of finding the particle at a given point behind the slits is also a cosine of the difference in the phase of the wavefunctions accumulated along the two paths from the source, one path passing through each slit.

\subsection{The phase difference from the action}

 At the output of the interferometer in Figure \ref{fig:interferometer} we want to determine the probability that the atom is observed to be in its excited state.  To do this we need to determine how the probabilities that the atom travels on either of the arms of the interferometer interfere with one another. We make a number of simplifying assumptions:  We will assume that the distance travelled by the atoms is small compared to the size of the source so that we can approximate the effects of the chameleon fifth force as a constant acceleration experienced by the atoms.
We write the constant acceleration as $a$, and choose our coordinates so that the acceleration  towards the source mass due to the chameleon, if it is present,  is in the $x$ direction, and this is also the direction of propagation of the laser beams.  The acceleration due to the Earth's gravity is $g$ and acts in the $z$ direction.
  With these assumptions the motion of the atoms is governed by  the Lagrangian
\begin{equation}
\mathcal{L} =\frac{m}{2}\dot{x}^2+ \frac{m}{2}\dot{z}^2 -max -mgz
\end{equation}
 A particle starts at $t=0$ and $x=z=0$ with initial velocity in the $z$ direction $\dot{z}=U$, it can then move on two possible paths, the $x$ component of which is   shown in Figure \ref{fig:interferometer}.  The equation describing these paths can be easily determined using the familiar formulae for motion in the presence of a constant acceleration.
\begin{itemize}
\item Path 1: At $t=0$ the particle receives an impulse that changes its  velocity in the $x$ direction by an amount $+V$.  It then moves freely  until time $T$ when is receives a second impulse that changes its velocity in the $x$ direction by an amount $-V$.  It then moves freely until time $2T$ at which point the measurement takes place.
When $0<t<T$
\begin{align}
x&= -\frac{1}{2}at^2 +Vt\\
z&= -\frac{1}{2}gt^2 +Ut
\end{align}
When $T<t<2T$
\begin{align}
x&= -\frac{1}{2}at^2 +VT\\
z&= -\frac{1}{2}gt^2 +Ut
\end{align}

Therefore the total action accumulated travelling along path 1 is
\begin{align}
S=&\frac{8}{3}m T^3 (g^2 + a^2) - 2mT^2 (2 gU +aV)+\frac{mT}{2}(2U^2 +V^2)
\end{align}

\item Path 2: The particle moves freely  until time $T$ when it receives an impulse that changes its velocity in the $x$ direction by an amount $+V$.  It then moves freely again until time $2T$ when it receives an impulse that changes its velocity in the $x$ direction by an amount $-V$, and the system is measured.
When $0<t<T$
\begin{align}
x&= -\frac{1}{2}at^2 \\
z&= -\frac{1}{2}gt^2 +Ut
\end{align}
When $T<t<2T$
\begin{align}
x&= -\frac{1}{2}at^2 +Vt-VT\\
z&= -\frac{1}{2}gt^2 +Ut
\end{align}

Therefore the total action accumulated travelling along path 2 is
\begin{align}
S=&\frac{8}{3}m T^3 (g^2 + a^2) - 2mT^2 (2 gU +aV)+\frac{mT}{2}(2U^2 +V^2)
\end{align}
\end{itemize}

We see that the action  is the same along both paths.  We recall that the probability of detecting the atoms at point 4, depends on the difference of the phases of the probability amplitude accumulated along these two paths.  These phases are proportional to the action integrated along each path.  But as we have found that the action is the same along both paths, these contributions cancel one another out.\footnote{   This happens because we have assumed that the particles are moving in a constant gravitational potential.  If we relaxed this assumption then a  difference in action between the two paths would be seen.}

\subsection{Phases from interacting with photons}
There is also a contribution to the phase of the atomic wave function due to the interaction with the laser beams.
The wave function describing the system consisting of an atom interacting with a photon must always be continuous.  A propagating photon is described by a wave equation, and so the phase of its wave function is proportional to $(i/\hbar) (\omega t - \vec{k}\cdot \vec{x})$, where $\omega$ is the energy of the photon, $\vec{k}$ is its momentum, with $\vec{x}$ and $t$ describing a point in space and time.  The electromagnetic vacuum, on the other hand, contains no propagating waves and so is described by a wave function which is constant and independent of time and space.  Therefore, to ensure continuity of the wavefunction of the combined system, when it absorbs a photon the wave function of an atom must change by a phase equal to that lost by the radiation field $(i/\hbar) (\omega T - \vec{k}\cdot \vec{X})$, where $\vec{X}$ and $T$ now describe the point in space and time at which this interaction took place.  

The wavefunction describing the atom at the time of measurement is a linear superposition of the two wavefunctions describing the evolution of the atom along the two paths of the interferometer.  This contains a piece which is the sum of the two sets of phases accumulated by the atom through its interactions with the laser beams. As the photon momentum is chosen to be only in the $x$ direction, the phase is independent of the motion of the atoms in the $z$ direction.
\begin{equation}
\psi(x_4,t_4)\propto e^{(i/\hbar)[\omega(t_4-t_3 +t_1)- k(x_4 -x_3 +x_1)]}+e^{(i/\hbar)[\omega t_2-k x_2]}
\label{eq:phase1}
\end{equation}
where the first term is the contribution from path 1 and the second term from path 2.
Therefore the probability of measuring the atom in its excited state at point 4, has a term proportional to 
\begin{equation} 
P\propto 2 \cos \frac{1}{\hbar}[\omega(t_4-t_3 +t_1)- k(x_4 -x_3 +x_1) -\omega t_2 +k x_2]
\label{eq:phase2} 
\end{equation}
We have chosen our coordinates such that $t_1=0$, $t_2=t_3 =T$ and $t_4 = 2T$, the corresponding expressions for the spatial positions of the interaction points can be determined from equations (\ref{eq:phase1})-(\ref{eq:phase2}), which results in
\begin{equation} 
P\propto 2 \cos \left[\frac{a T^2 k}{\hbar}\right]
\end{equation}

We have shown that the probability that the atom is observed in its excited state at point 4 is a function of the local acceleration $a$.  This is why atom interferometry works as a technique for measuring small accelerations.  In practice the experiment is performed, not with single atoms, but with a large cloud of atoms.  Therefore a large number of measurements can be accumulated rapidly, and $P$ can be determined to a high degree of precision.

\section{Atom interferometry searches for the chameleon}
Atom interferometry can be used to measure any small acceleration that occurs in the direction of propagation of the laser beams.  
The technique of atom interferometry does not care about the origin of the accelerating force, it could be due to gravity or to a novel fifth force.  Indeed one practical application of atom interferometry that is under development is in the guidance systems of submarines.  If the interferometry apparatus can be made sufficiently small, then it could be placed aboard the submarine, and the acceleration of the submarine could be tracked, and after integrating twice the position of the submarine could be determined with out any need for the submarine to signal to the outside world \cite{4570035}.

In order to use atom interferometry for less militaristic purposes, we need to think carefully about how to design an experiment in order to make it suitable for searching for chameleon fields. 
When trying to measure gravity with an interferometry experiment, we have a very large source close to hand; the Earth.  The earth's gravitational field easily penetrates into the interior of a vacuum chamber and can be detected using atom interferometry as we have described above. However a chameleon field due to the Earth would not penetrate into the vacuum chamber in the same way, precisely because of its chameleonic nature - recall equation (\ref{eq:mmin}).  

The Compton wavelength of an interaction is proportional to $1/m_{\rm bg}$, where $m_{\rm bg}$ is the mass of the particle transmitting the force.  It determines the maximum distance over which an interaction can propagate (more precisely, the effects of the force are exponentially suppressed on distance scales larger than the Compton wavelength).  This is why electromagnetic interactions, transmitted by the massless photon, can propagate over arbitrarily large distances, whilst the weak force, transmitted by the heavy $W$ and $Z$ bosons, is confined to short distance scales. As we showed in equation (\ref{eq:mmin}) the chameleon becomes massive in a dense environment, and this means that chameleon particles cannot transmit a force very far through a dense environment.  As we have seen this is the source of chameleon screening, but it also means that chameleon forces from outside the vacuum chamber cannot be transmitted into the interior, assuming the walls of the vacuum chamber are at least $\sim 1 \mbox{ mm}$ thick. 

As a result,  inside the vacuum chamber the only forces acting on the atom are the gravitational field due to the Earth, and the gravitational and chameleon attractions to the source object. However the experiment is only sensitive to the component of the forces parallel to the direction of propagation of the laser beams.  Therefore if we orientate our experiment as shown in Figure \ref{fig:drawing}, so that the atoms are held adjacent to the source mass, and the laser beams propagate perpendicular to the direction of the gravitational field due to the Earth, then the outcome of the interferometry experiment will only be affected by the source mass and the gravitational field due to the Earth plays no role.

\begin{figure}[tp]
\centering
\includegraphics[]{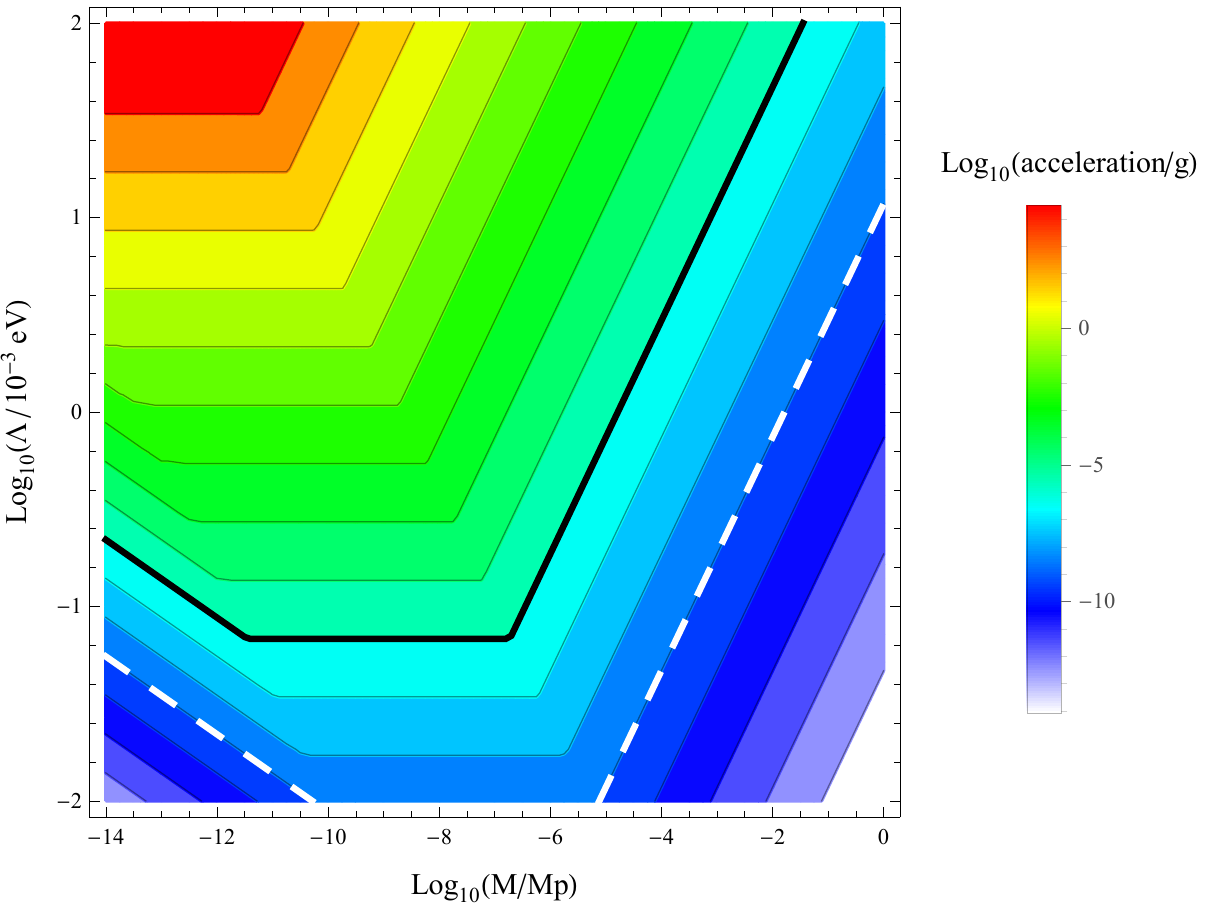}
\caption{Contour plot showing acceleration of rubidium atoms (normalised to the acceleration $g$ of free fall on earth) due to the chameleon force outside a sphere of radius $R_A=1\mbox{\,cm}$  and screening factor $\lambda_A=\frac{3 M_A \phi_{\rm bg}}{\rho_A R_A^2} $. The atom and sphere are placed centrally in a $10\mbox{\,cm}$-radius vacuum chamber containing $10^{-10}\mbox{\,Torr}$ of hydrogen. The $\Lambda-M$ area above the heavy solid black line was recently excluded by a first atom interferometer experiment measuring $10^{-6}g$ \cite{Hamilton:2015zga}. With modest attention to systematic errors this can move down to the heavy dashed white line.  For $\Lambda \ge 10\,\mbox{meV}$, atom interferometry could sense chameleon physics up to the Planck mass  $M_P$. }
\label{fig:exclusion}
\end{figure}

In \cite{Burrage:2014oza}, with E. A. Hinds we proposed that an appropriate experiment to search for chameleon forces could be performed using Raman interferometry (where two photons are used to excite the atom with a frequency difference tuned to match the difference in energy levels) and Rubidium atoms.  We suggested that accelerations down to $10^{-6}g$ could be detected with a preliminary experiment, and that with attention to systematics precision of $10^{-9}g$ could be achieved with current technology.  Remarkably, within a few months of our proposal, in \cite{Hamilton:2015zga} such an experiment was performed. Using a caesium matter-wave interferometer they did indeed achieve a sensitivity of $10^{-6}g$.   The constraints this places on the chameleon are shown in Figure \ref{fig:exclusion}, along with the forecasted exclusion for a future upgraded version of this experiment. 

\section{Future prospects}
We would like to be able to perform an experiment, or series of experiments, which would cover the whole of the chameleon parameter space.  This would allow us to either definitively detect or exclude dark energy models that screen through the chameleon mechanism.  One way to improve the sensitivity of the experiment could be by making a clever choice of the shape of the source mass used in the experiment.  Unlike the gravitational force, which largely doesn't care about the shape of a source only its mass, the strength of the chameleon force depends on the shape chosen for the source object.  This is because of the non-linearities  mentioned earlier that are an intrinsic part of the theory.  Unfortunately these non-linearities make the behaviour of the chameleon field very difficult to calculate, for anything other than the spherical sources discussed above.  Recently, however, it has been shown  \cite{Burrage:2014daa} that when a spherical source of the chameleon field is a strongly perturbing object, by deforming this source from a sphere to an ellipse we can increase the strength of the chameleon force by up to 40\%.  Investigation of other possible source shapes is currently underway.  If it is possible to mill, or perhaps 3D print, a source mass of the optimum shape to source the chameleon field
it may finally be possible to complete the search for the chameleon, and possibly detect the dark energy particle in the laboratory.

\bibliographystyle{JHEP}
\bibliography{arXiv_version}

\end{document}